%% file: main.tex
\documentclass[runningheads]{llncs}
\usepackage{graphicx}
\usepackage{paralist}
\usepackage{enumitem}
\usepackage{listings}
\usepackage{color}
\usepackage{xcolor}
\usepackage{textcomp}
\usepackage{setspace}
\usepackage{hyperref}
\usepackage{url}
\usepackage{longtable}

\makeatletter
\newcommand{\setword}[2]{%
  \phantomsection
  #1\def\@currentlabel{\unexpanded{#1}}\label{#2}%
}
\makeatother

\usepackage{soul}

\newcommand{\comment}[1]{}

\usepackage{quoting}
\quotingsetup{leftmargin=0.05\linewidth, rightmargin=0.05\linewidth, indentfirst=false, vskip=3pt}
  
\usepackage{subcaption}

\begin{document}
%
%\title{Turning Transport Data into EU Compliance while Enabling a Multimodal Transport Knowledge Graph}
\title{Turning Transport Data to Comply with EU Standards while Enabling a Multimodal Transport Knowledge Graph}
\titlerunning{Turning Transport Data to Comply with EU Standards}
\author{Mario Scrocca\orcidID{0000-0002-8235-7331} \and
Marco Comerio\orcidID{0000-0003-3494-9516} \and
Alessio Carenini\orcidID{0000-0003-1948-807X} \and
Irene Celino\orcidID{0000-0001-9962-7193}
}
\authorrunning{M. Scrocca, M. Comerio, A. Carenini and I. Celino}
\institute{Cefriel -- Politecnico of Milano \\Viale Sarca 226, 20126 Milano, Italy
%\institute{Cefriel, Milan, Italy
\email{name.surname@cefriel.com}}
\maketitle % typeset the header of the contribution
\begin{abstract}
Complying with the EU Regulation on multimodal transportation services requires sharing data on the National Access Points in one of the standards (e.g., NeTEx and SIRI) indicated by the European Commission. These standards are complex and of limited practical adoption. This means that datasets are natively expressed in other formats and require a data translation process for full compliance. 

This paper describes the solution to turn the authoritative data of three different transport stakeholders from Italy and Spain into a format compliant with EU standards by means of Semantic Web technologies. Our solution addresses the challenge and also contributes to build a multi-modal transport Knowledge Graph of interlinked and interoperable information that enables intelligent querying and exploration, as well as facilitates the design of added-value services.

\keywords{Transport Data \and Semantic Data Conversion \and Multimodal Transport Knowledge Graph \and Transport EU Regulation}
\end{abstract}
\input{sections/introduction}
\input{sections/challenges}

\input{sections/solution}

\input{sections/evaluation}
\input{sections/conclusion}

\footnotesize
\subsubsection*{Acknowledgments}
The presented research  was partially supported by the SPRINT project  (Grant Agreement 826172),  co-funded  by  the European  Commission  under  the  Horizon  2020  Framework Programme and by the SNAP project (Activity Id 19281) co-funded by EIT Digital in the Digital Cities Action Line.

\bibliographystyle{splncs04}
\bibliography{biblio}

\end{document}

%% file: sections/introduction.tex
\section{Introduction}\label{sec:intro}

% Da proposte talk: EU missions, Regulation 2017/1926, definition of the problem (data compliance with EU standards), brief description of SNAP. Here or in Section 2: results of SNAP survey. Dominio transporto caratterizzato da eterogeneità. NeTeX (lo standard scelto) non è molto usato.

Semantic interoperability in the transportation sector is one of the European Commission challenges: establishing an interoperability framework enables European transport industry players to make their business applications ‘interoperate’ and provides the travelers with a new seamless travel experience, accessing a complete multi-modal travel offer which connects the first and last mile to long distance journeys exploiting different transport modes (bus, train, etc.). 

With the ultimate goal of enabling the provision of multi-modal transportation services, the EU Regulation 2017/1926\footnote{EU Reg. 2017/1926, cf. \url{https://eur-lex.europa.eu/eli/reg\_del/2017/1926/oj}} is requiring transport service pro\-vid\-ers (i.e., transport authorities, operators and infrastructure managers) to give access to their data in specific data formats (i.e., NeTEx\footnote{NeTEx, cf. \url{http://netex-cen.eu/}} and SIRI\footnote{SIRI, cf. \url{http://www.transmodel-cen.eu/standards/siri/}}) through the establishment of the so-called National Access Points (NAP). 

A survey conducted by the SNAP project\footnote{SNAP (Seamless exchange of multi-modal transport data for transition to National Access Points), cf. \url{https://www.snap-project.eu}} revealed that transport service providers have a poor knowledge of the EU Regulation 2017/1926 and its requirements  and they do not use or even know the requested standards (NeTEx and SIRI). The transport stakeholders deem the conversion of their data to these standards as technically complex: they would need to dedicate a notable amount of resources to this effort, but often they lack such resources. 

% REPHRASED Within the SNAP project, we designed and developed an innovative solution for data conversion, decreasing the time required to perform it and hiding its complexity. Our solution, described in this paper, is based on Semantic Web technologies and enables the conversion of complex transportation data into EU-mandated standards: it fit the needs of transport service providers to render their legacy data interoperable and compliant with the standards.

We designed and developed an innovative solution for data conversion, based on Semantic Web technologies, hiding the complexity of the conversion and enabling flexibility to address different scenarios and requirements. Our solution, described in this paper, has been validated within the SNAP project to enable the conversion of complex transportation data into EU-mandated standards. The proposed approach fits the needs of transport service providers rendering their legacy data interoperable and compliant with the regulation. Moreover, enabling a multi-modal transport knowledge graph, it also fosters data harmonization for the design and development of added-value travel services. 

This paper is organized as follows: Section~\ref{sec:challenges} presents our vision on how to address the transport data conversion using Semantic Web technologies; Section~\ref{sec:chimera} describes our solution; Section~\ref{sec:pilots} provides technical and business evaluations; Section~\ref{sec:conclusion} draws conclusions and ongoing works.
% and Section~\ref{sec:lessons}

%% file: sections/challenges.tex
\section{Challenges and Vision}\label{sec:challenges}

% Solve the problem applying semantic web technologies. Why not an in-house custom solution (often used in transport domain)?
% Comparison w.r.t. SOTA: only the classification of the approach as “any-to-one centralized” available (see D01). TBD. approcci mapping semantico??? (da paper David? Da D5.3?) + metodologia Sequeda (descrizione anche parte business?) [?] Slide produced by Irene for POLIS “A few more details on the SNAP Solution”}

Complying with the EU Regulation on multi-modal transportation services requires sharing data on the National Access Points in one of the standards indicated by the European Commission. This means that each affected organization -- transport authority, transport operator and infrastructure manager -- has to produce data in such formats. The goal of the European approach is clear and praiseworthy: limit the heterogeneity of data formats, require specific levels of quality and expressivity of data, and pave the way for interoperability.

As mentioned above, the mandated standards are complex and of limited practical adoption. NeTEx is a very broad and articulated XML schema and it was created as an interchange format rather than an operational format. Moreover, very few production system adopt it, so converting to NeTEx requires a deep understanding of its intricacies. NeTEx contains 372 XSD files with 2898 complex types, 7134 elements and 1377 attributes (without considering the national profiles that bring additional country-specific content). 

This means that today existing datasets are natively expressed in other formats and require a data translation process for full compliance. Point-to-point conversion is of course possible, but it requires a complete knowledge of the target formats and the correspondences between the original schemata and the target standards. Our approach hides this complexity from the stakeholders, letting them keep using their current legacy systems.

The transportation domain is characterised by a proliferation of formats and standards which do not facilitate system interoperability. However, there are efforts that try to overcome the incompatibility challenge, such as Transmodel\footnote{Transmodel, cf. \url{http://www.transmodel-cen.eu/}}, which aims at representing a global conceptual model for this domain. Indeed, Transmodel is the rich and comprehensive reference schema (its data dictionary only contains 1068 concepts) on which other specific formats are based, such as NeTEx and SIRI themselves.

This is clearly an opportunity for Semantic Web technologies and for the adoption of an any-to-one centralized mapping approach \cite{vetere2005}, i.e., a global conceptual model (i.e. an ontological version of Transmodel) allowing for bilateral mappings between specific formats and the reference ontology \cite{comerio2019turn}. 

\begin{figure}[t!]
  \centering
\begin{subfigure}{.45\textwidth}
  \centering
  \includegraphics[width=.6\textwidth]{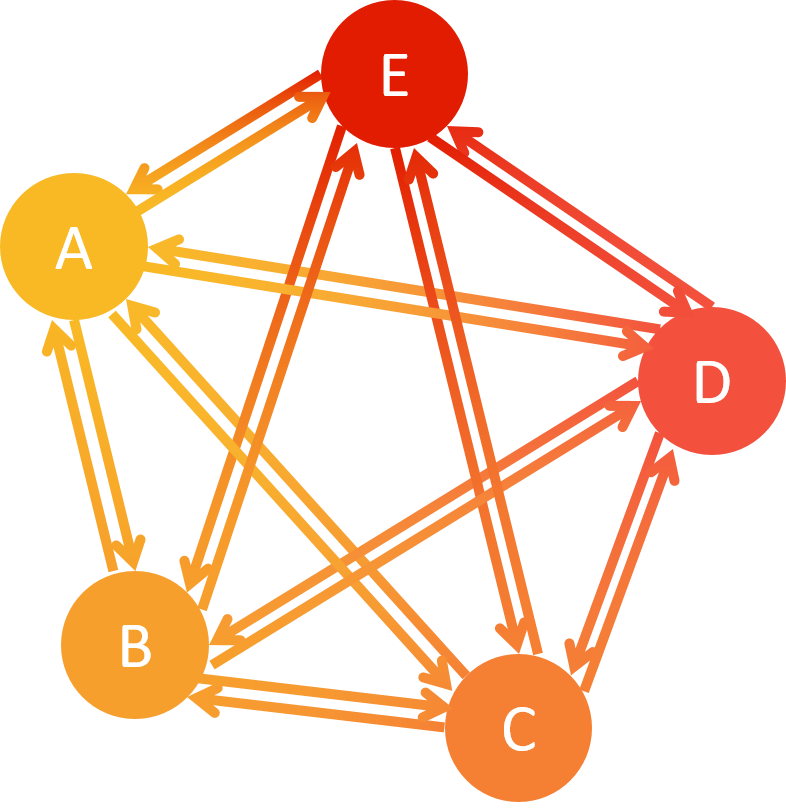}
  \label{fig:sub-first}
\end{subfigure}
\begin{subfigure}{.45\textwidth}
  \centering
  \includegraphics[width=.6\textwidth]{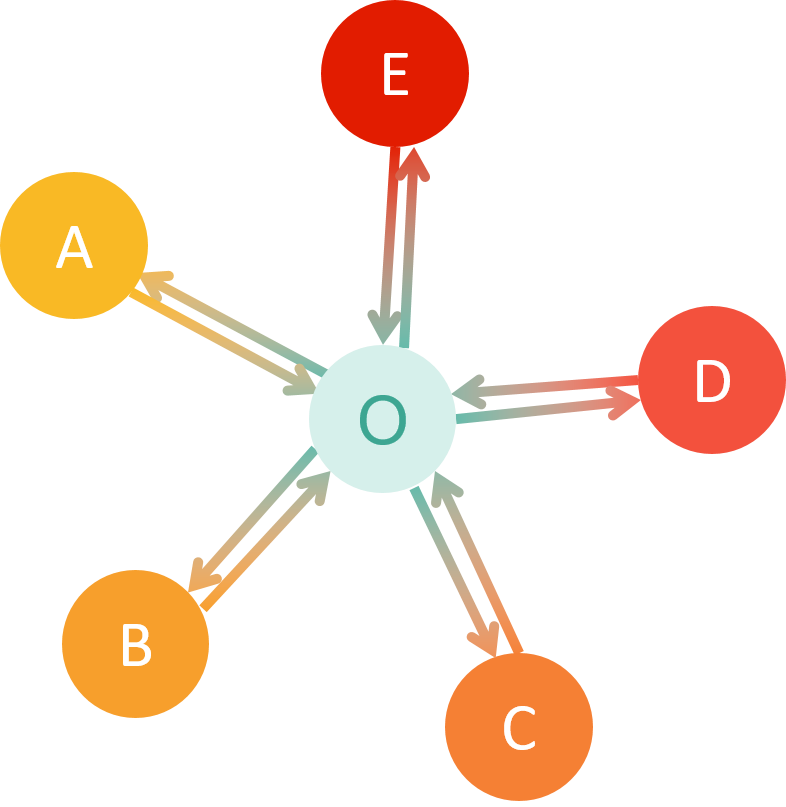}  
  \label{fig:sub-second}
\end{subfigure}
\caption{Mappings required without and with a global conceptual model}
\label{fig:num-mapping}
\end{figure}

This approach provides a twofold advantage.
On the one hand, it reduces the number of  mappings required for conversion and compliance, making the management of complexity manageable: if there are n different formats, the number of mappings is $2n$ instead of $2n(n-1)$ (cf. Figure~\ref{fig:num-mapping}).

\begin{figure}[b!]
    \centering
    \includegraphics[width=.78\textwidth]{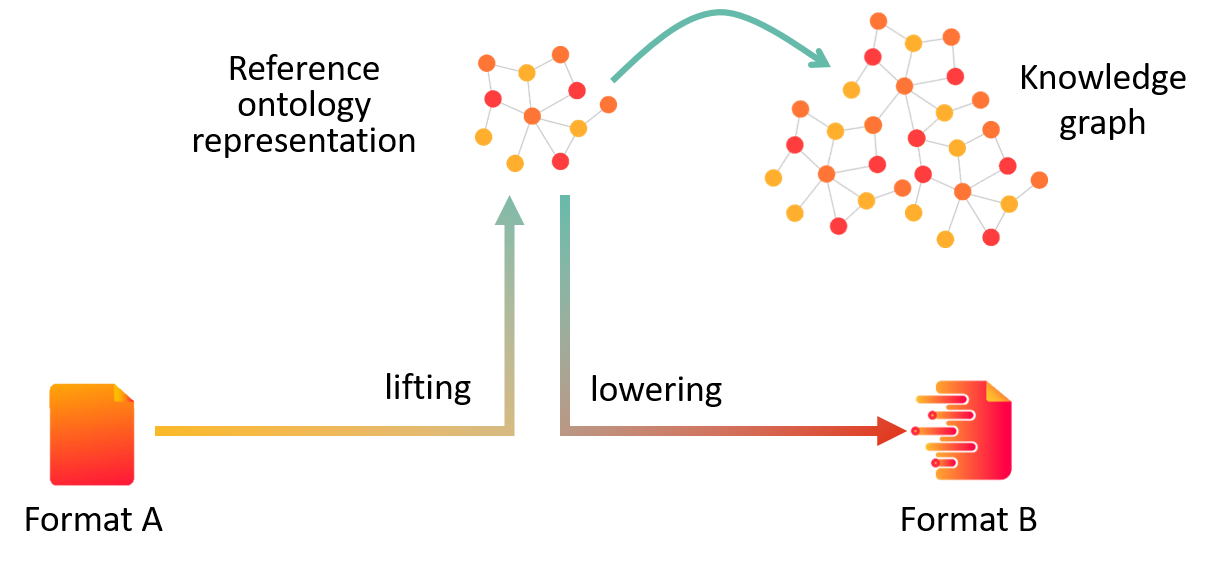}
    \caption{Building a knowledge graph as a ``collateral effect" of data conversion}
    \label{fig:collateral}
\end{figure}

On the other hand, the two-step conversion process (i.e., lifting from the original format to the reference ontology and then lowering from the ontological version to the target format) has a positive ``collateral effect'': once lifted to the semantic conceptual level, data contributes to build a multi-modal transport knowledge graph of interlinked and interoperable information (cf. Figure \ref{fig:collateral}).

Taking the example of static transport information (e.g., timetables of public transportation), the target format required by the EU Regulation is the already mentioned NeTEx. Even when all data from different transport stakeholders are converted to NeTEx and shared on a National Access Point, still the challenges of data integration and reuse are not easily solved. However, the availability of a Semantic Web-enabled knowledge representation level, with data expressed with respect to a reference ontology, allows for seamless integration through automatic linking, for intelligent querying and exploration, and for the facilitation of added-value service design. Examples of such services are Intelligent Transport Systems -- with new features for travelers such as planning of door-to-door journeys through the usage of multiple transportation modes -- or Decision Support Systems for government bodies e.g., for urban planners, providing statistics on mobility usage and supporting the analysis of territory coverage, to comply with the European Green Deal.

Moreover, this multi-modal transport knowledge graph can be implemented with an incremental approach, as suggested by \cite{sequeda2019pay}, with the possibility to add additional data from multiple providers at different times, still highlighting the advantages and showcasing the opportunities of such a method.

In a nutshell, while addressing the challenge of compliance with the EU Regulation on transport information sharing, we advocate the adoption of a Semantic Web approach that not only solves the issue of data conversion, but also provides a strong and solid solution for data interoperability and for the creation of a new generation of Intelligent Transport Systems, which support the European Commission vision on seamless travel experience for travellers.

%% file: sections/solution.tex
\section{Our Solution}\label{sec:chimera}

% From D01. My suggestion: “SNAP as a whole” + “in this paper, we focus on Data Conversion that adopts SWT” + from D01 “The Chimera Converter” + examples from “SNAP in Action” + D5.2. Only ref/brief description of the Transmodel Ontology

In this section, we show how we addressed the challenges and implemented the vision of Section \ref{sec:challenges}. %in the context of the SNAP project, that solves the problem to turn transport data into compliance with the EU Regulation 2017/1926. 
In particular, we designed a solution that implements the any-to-one centralized approach to semantic interoperability, with a global conceptual model %(i.e., a preliminary ontological version of Transmodel) 
to allow for bilateral mappings between specific formats and the reference ontology. %Our solution has been tested to turn static transportation data into the standard NeTEx, covering the cases in which the source data are: (i) GTFS files including static data describing public transportation services, (ii) JSON files including data in proprietary format addressing the description of airport facilities.
After an overview on similar approaches and related efforts, in this section we explain the requirements, the architecture and the technical choices of Chimera, our open source framework to address the conversion challenge.

%%%%%%%%%%%%%%%%%%%%%%%%%%%%%%%%%%%%%%%%%%%%%%%%%%%%%%%%%%%%%%%%%%%%%%%%%%%%%%%%%%%%
\subsection{Related Works}\label{sub:related}
Data transformations based on a common data model, such as the ones described in the previous section, require implementing two different processes. The former, which in the context of semantic technologies have been named ``lifting", extracts knowledge from the data source(s), and represents it according to the common data model. The latter, which has been named as ``lowering", accesses such information and represents it according to a different data model and serialisation format. 

Different approaches have been proposed over the years to deal with data lifting to RDF. Most of them rely on code or format-specific transformation languages (such as XSLT applied to XML documents). Declarative languages have nonetheless been specified to deal with specific data sources, such as relational databases (R2RML~\cite{r2rml}), XML documents (xSPARQL~\cite{bischof2012mapping}) or tabular/CSV data (Tarql~\cite{cyganiak2015tarql}). Other languages instead allow creating data transformations that extract information from multiple data sources, such as RML~\cite{dimou2014rml} (a generalisation of the R2RML language) and SPARQL-Generate~\cite{lefrancois_ekaw_2016}. All the cited approaches and languages support the representation of the source information according to a different structure, and the modification of values of specific attributes to adapt them to the desired data model.
%\textcolor{red}{Another way to achieve lifting is to use an ORM-like approach, using marshalling/unmarshalling libraries to obtain an in-memory representation of the incoming data as objects, and then exploiting annotations stating how to map each class and attribute to which ontological class or property. This approach is implemented in RDFBeans\footnote{RDFBeans, cf.~ \url{https://github.com/cyberborean/rdfbeans}}, Empire\footnote{Empire, cf.~ \url{https://github.com/mhgrove/Empire}}, and has been studied during the ST4RT project\footnote{ST4RT project homepage, cf.~ \url{http://www.st4rt.eu/}} \cite{carenini2018st4rt}. While it allows using classical object-oriented programming techniques, its main drawback is memory consumption, as all the information from the incoming data source must be kept in memory to be used, and it is therefore ill-suited for large input documents.}

Data lifting using semantic technologies has been also streamlined inside different semantic-based ETL (``Extract, Transform and Load") tools. UnifiedViews~\cite{knap2014unifiedviews} and LinkedPipes~\cite{klimek2016linkedpipes} have been implemented during the years, providing environments fully based on Semantic Web principles to feed and curate RDF knowledge bases. A different approach is used by Talend4SW\footnote{Talend4SW, cf.~\url{https://github.com/fbelleau/talend4sw}.}, whose aim is to complement an already existing tool (Talend) with the components required to interact with RDF data. An ETL process is based not only on transformation steps. Other components, such as message filtering or routing, are usually involved. The main categorisation of the components and techniques that can be used in an integration process is the Enterprise Integration Patterns~\cite{hohpe2004enterprise}, which influenced heavily our work (cf. Section~\ref{sub:arch}).

At the opposite side of the transformation process, lowering has received less attention from the academic community, and there is not a generic declarative language to lower RDF to any format. xSPARQL~\cite{bischof2012mapping} provides a lowering solution specifically for XML, letting developers embed SPARQL inside XQuery. 
%Another way to achieve lowering is provided by ORM-like solutions, which also offer the possibility to perform data lifting. Exploiting annotations attached to object-oriented programming classes, those solutions can extract data from an RDF graph and create in-memory instances of objects, which can then be marshalled to the desired format with other libraries. Again, the main drawback of annotations-based solutions is memory consumption, because before performing marshalling all the information must be available as objects in memory.
A downlift approach was proposed in~\cite{debruyne2017lightweight}, by querying lifting mappings to recreate a target CSV format. In our work, we adopt a declarative approach to implement lowering to XML, relying on SPARQL for querying data and on template engines to efficiently serialise data according to the target format.

Addressing both lifting and lowering is also possible by adopting an object-relational mapping approach (ORM), using and object-oriented representation of both the ontological and non-ontological resources through annotations and mashalling and unmashalling libraries. This approach is implemented in RDFBeans\footnote{RDFBeans, cf.~ \url{https://github.com/cyberborean/rdfbeans}} and Empire\footnote{Empire, cf.~ \url{https://github.com/mhgrove/Empire}} and we also applied it in the past~\cite{carenini2018st4rt2}. However, its main drawback is memory consumption, making it unsuitable for large data transformation; this is why we turned to the approach described in this paper.

The specific problem of producing datasets complying to the  EU Regulation 2017/1926, converting GTFS datasets into NeTEx, is currently being addressed by means of a non-semantic approach by Chouette~\cite{gendre37chouette}. Even if the implementation does not use Semantic Web languages, the conversion process is implemented in the same way. In Chouette, GTFS data is first loaded inside a relational database strongly based on the Neptune data model, and users can then export data from the database to NeTEx. Such database and related data model perform the same role than RDF and the Transmodel ontology in our approach. Chouette, as well as any direct translation solution, can address and solve the conversion problem; however, it does not build a knowledge graph nor it helps in enabling an easy integration or further exploitation of the converted data.

%%%%%%%%%%%%%%%%%%%%%%%%%%%%%%%%%%%%%%%%%%%%%%%%%%%%%%%%%%%%%%%%%%%%%%%%%%%%%%%%%%%%
\subsection{Chimera requirements}\label{sub:req}
The design of our software framework -- that we named Chimera -- aimed to address the data conversion challenge highlighted in Section~\ref{sec:challenges} and, in the meantime, to offer a generic solution for data transformation adoptable in domains different than the transportation one. 
The requirements that guided our work can be summarized as follows:
%\begin{itemize}
%(1) data conversion with semantics
    %\item 
    (i) the solution should address the conversion challenge following the any-to-one centralized mapping approach \cite{vetere2005}, by employing Semantic Web technologies;
%(2) applicable to both datasets and messages
    %\item 
    (ii) the conversion should support two different data transformation scenarios: batch conversion (i.e., conversion of an entire dataset, typical of static data) and message-to-message mediation (i.e., translation of message exchanges in a service-centric scenario, typical of dynamic data);
%(3) reusable code + configuration
    %\item 
    (iii) the solution should minimise the effort required to the adopters to customize the data conversion process. 
%\end{itemize}

%%%%%%%%%%%%%%%%%%%%%%%%%%%%%%%%%%%%%%%%%%%%%%%%%%%%%%%%%%%%%%%%%%%%%%%%%%%%%%%%%%%%
\subsection{Chimera architecture}\label{sub:arch}
%(1) generic pipeline to address the conversion
The design of the Chimera architecture follows a modular approach to favour the conversion customization to different scenarios and formats, thus guaranteeing the solution flexibility. 
The decision of the modular approach is based on the assumption that a data conversion process could be broken down in a pipeline of smaller, composable and reusable elements. 

%(3) inspired by EIT with ref to related
We took inspiration from the already mentioned Enterprise Integration Patterns (EIP) approach, that offers best practices and patterns to break down a data processing flow into a set of building blocks. In this sense, Chimera is similar to the other data transformation solutions mentioned in Section~\ref{sub:related}.
With respect to EIP terminology, a Chimera converter implements a \emph{Message Translator} system, i.e., a pattern to translate one data format into another; in particular, a Chimera generic pipeline can be seen as a composition of specialized transformers called \emph{Content Enricher}. %and accessing an RDF Graph shared among the different blocks\mario{Provato a rifrasare, Semantic Data Transformation Pipelines ce lo siamo inventati noi}.

\begin{figure}[t!]
  \centering
  \includegraphics[width=\linewidth]{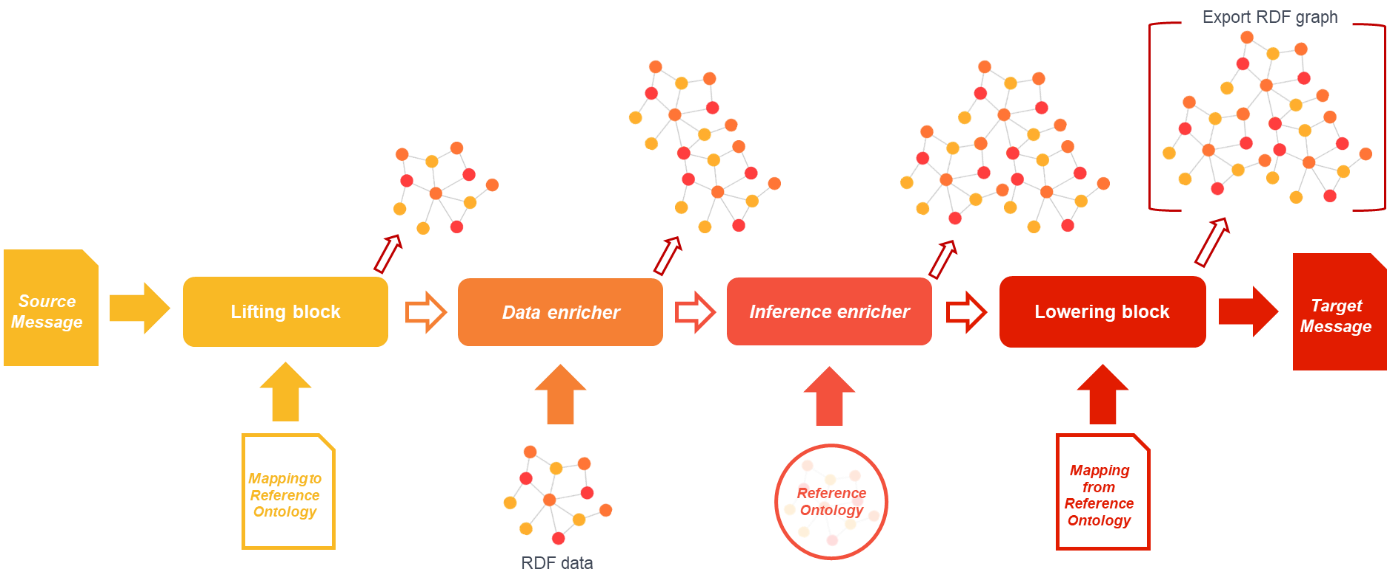}
  \caption{A data conversion pipeline supported by Chimera.}
    \label{fig:pipeline}
\end{figure}

%(2) basic + advanced blocks
The generic pipeline for a data conversion process is illustrated in Figure~\ref{fig:pipeline}. The basic conversion includes the already mentioned lifting and lowering elements; however, to support a wider set of scenarios, the Chimera architecture includes additional building blocks. 

Therefore, in the design of the Chimera architecture we identified the following typologies of building blocks:
\begin{itemize}
\item \textbf{Lifting block} (lifting procedure): this block takes structured data as input and, by means of mappings between the source data model and the reference ontology, transforms it into its ontological representation (an RDF graph);
\item \textbf{Data enricher block} (merging/linking procedure): this (optional) block takes a set of additional RDF graphs and integrates them within the current graph; this block can be useful when additional data sources, already represented using the reference ontology, are available to enrich the current knowledge graph (i.e., information not present in input messages but needed for the conversion); %Add case CONSTRUCT query to enrich the graph?
\item \textbf{Inference enricher block} (inference procedure): this (optional) block takes the RDF graph and, based on a set of ontologies (containing axioms or rules), enables an inference procedure to enrich the graph with the inferred information;
\item \textbf{Lowering block} (lowering procedure): this block queries the RDF graph and, by means of mappings between the reference ontology and the target data format, transforms the data into the desired output format.
\end{itemize}

In the conversion pipeline, the RDF graph is an intermediate product and it is not a necessary output of the data transformation, thus it could be discarded once the pipeline execution is completed. However, we believe that this ``collateral effect" of building an RDF graph is one of the strengths of our approach, because it enables the incremental building of a knowledge graph. In our multi-modal transport scenario, this is of paramount importance, because it allows to build a knowledge graph covering different modes and therefore representing a precious source for multi-modal intelligent transport systems. The Chimera pipelines, therefore, can be configured to save the enriched RDF graph as an additional output of the transformation.

%%%%%%%%%%%%%%%%%%%%%%%%%%%%%%%%%%%%%%%%%%%%%%%%%%%%%%%%%%%%%%%%%%%%%%%%%%%%%%%%%%%%
\subsection{Chimera implementation}\label{sub:impl}
%(1) based on Camel, inheriting its advantages
We implemented Chimera on top of Apache Camel\footnote{Apache Camel, cf. \url{https://camel.apache.org/}.}, a Java integration framework to integrate various systems consuming or producing data. We chose Camel to inherit its features and advantages: not only it is a completely open-source, configurable and extensible solution, but it also implements best practices and patterns to solve the most common integration problems, including the already-mentioned Enterprise Integration Patterns; finally, being a robust and stable project, Camel supports out-of-the-box several components, runtimes and formats to access and integrate a large set of existing system and environments. 

Chimera implements data conversion by exploiting Camel's \emph{Routes} to process data, i.e., custom pipelines with different components to enable a specific transformation. Those Routes can be defined programmatically or, exploiting the integration with Spring\footnote{Spring, cf. \url{https://spring.io/}.}, they can be simply configured through an XML file.
Thanks to the huge set of Camel's pre-defined components, Chimera can achieve both batch conversion and message-to-message mediation, and it can leverage multiple input and output channels: data can be acquired by polling directories or FTP servers, or by exposing REST services, Web APIs and SOAP services, by using different publish-subscribe technologies such as JMS or AMQP. 

%(2) custom block implementation with ref to libraries
The Chimera framework implements the specific blocks illustrated in Section~\ref{sub:arch} with the help of Semantic Web technologies. On top of Chimera implementation, pipelines can be defined by simply providing a configuration file, without the need to add any code. 

The basic idea of Chimera is to define a pipeline composed of different blocks and to ``attach" an RDF graph to the specific ``message" going through the Camel route. In this way, each block can process the incoming data by relying on a shared global RDF graph. All Chimera components use the RDF4J library\footnote{RDF4J, cf. \url{https://rdf4j.org/}} to process and handle RDF data and to operate on a Repository interface that can be in-memory, local or a proxy to a remote triplestore.

In the current release of Chimera, we provide the following blocks:
\begin{itemize}
    \item \textbf{RMLProcessor}: a \emph{lifting block} based on RML~\cite{dimou2014rml}, exploiting our fork\footnote{Cf.  \url{https://github.com/cefriel/rmlmapper-cefriel}.} of the rml-mapper library\footnote{Cf. \url{https://github.com/RMLio/rmlmapper-java}.}; our modified version of the mapper extends RML to declare a set of InputStreams (i.e. a generic source stream rather than individual files\footnote{We borrowed the idea from the Carml implementation of RML, cf. \url{https://github.com/carml/carml\#input-stream-extension}.}) as logical sources in the RML mapping file.
    \item \textbf{TemplateProcessor}: a \emph{lowering block} based on Apache Velocity\footnote{Cf. \url{https://velocity.apache.org/}.} to implement a template-based solution to query the RDF graph, process the result set and generate the output data in the desired format\footnote{We implemented this approach both as a Chimera block and as a standalone tool available at \url{https://github.com/cefriel/rdf-lowerer}.}. 
    %\item \textbf{ST4RTLiftingProcessor} and \emph{ST4RTLoweringProcessor}: a lifting and lowering block based on the annotations-based method developed in the ST4RT project.
    \item \textbf{DataEnricherProcessor}: a \emph{data enricher block} to add a set of RDF sources to the pipeline graph.
    \item \textbf{InferenceEnricherProcessor}: an \emph{inference enricher block} loading a set of ontologies to enrich the pipeline graph with inferred information; it exploits the inferencing capabilities and configurations of the repository\footnote{Cf. for instance RDFS inference on in-memory/native RDF4J stores \url{https://rdf4j.org/documentation/programming/repository/\#rdf-schema-inferencing}.}.
    \item A set of utility processors (\textbf{AttachGraph}, \textbf{DumpGraph}) to handle the initialization and export of the RDF graph.
\end{itemize}

%(3) open source with sample pipeline
Chimera is released as an open source software framework, with an Apache 2.0 license, and is available at \url{https://github.com/cefriel/chimera}.
An example pipeline showing how to configure a semantic conversion pipeline using the described lifting, lowering and data enricher blocks is also available\footnote{Cf. \url{https://github.com/cefriel/chimera/tree/master/chimera-example}.} and can be tested by following the instructions of the README file.

%% file: sections/evaluation.tex
\section{Real-world Evaluation of our Solution}\label{sec:pilots}

In this section, we describe the activities performed within the SNAP project to assess the proposed approach and the Chimera converter on concrete scenarios and real datasets. In Section~\ref{sec:snap}, we describe the pilot scenarios, the involved stakeholders and the differences among the use cases. Then, we provide an evaluation of our proposed solution on four main dimensions:
\begin{itemize}
    \item \emph{Mappings}: in Section~\ref{sec:mappings} we discuss the mappings' definition procedure, presenting a conceptualization of the different steps required to enable a conversion pipeline, and an assessment considering the pilots scenarios.
    \item  \emph{Flexibility}: in Section~\ref{sec:pipeline} we discuss the flexibility provided by Chimera in composing customized pipelines to address the specific requirements of the different pilots.
    \item \emph{Performance}: in Section~\ref{sec:performances} we present the actual performance results and comment on conversion pipelines' execution using different input datasets.
    \item \emph{Business Viability}: in Section~\ref{sec:business} we present considerations emerged from pilots on the business viability of the proposed solution.
\end{itemize}

\subsection{Evaluating the SNAP solution}\label{sec:snap}

% ***** Questa parte forse alla fine la escluderei ******** %
%The SNAP proposition used for communication and promotion have been crafted to promote the characteristics of the solution in addressing the larger need for interoperability in the transportation industry. The SNAP solution is promoted as:
%\begin{inparaenum}[(i)]
% \item \emph{Unique}: no proven, market-ready solutions cover the entire data conversion process enabled by SNAP; 
%\item \emph{Innovative}: SNAP provides data conversion based on Semantic Web technologies; 
%\item \emph{Simple}: SNAP hides the complexity of the conversion process and of target standards;
%\item \emph{Timesaving}: SNAP prevents acquiring the knowledge or contracting the competence to implement a custom solution.   
%\end{inparaenum}

To evaluate our solution on both a technical and business side, in the context of the SNAP project we identified a set of stakeholders in the transport domain affected by the regulation and necessitating to convert their data. The identified pilots allowed us to test both functional and non-functional requirements of our conversion solution and let us generate valuable outputs for the involved stakeholders that not only obtained data compliant with the regulation, but they also received the RDF representation of their data, with respect to the common reference ontology based on Transmodel.

Three different pilots, covering the region of Madrid, the city of Milan, and the city of Genova, were executed. In each region, we engaged with the corresponding stakeholders to gather uses cases involving different actors identified by the EU regulation, i.e., transport operators, transport authorities and infrastructure managers. 

In the region of Madrid, our \emph{transport authority} pilot, we transformed into NeTEx the GTFS data sources of both the Consorcio Regional de Transportes de Madrid\footnote{CRTM, cf. \url{https://www.crtm.es/}} (Light Railway, Metro Madrid, Regional Railway) and EMT Madrid\footnote{EMT Madrid, cf. \url{https://www.emtmadrid.es/}} (the public bus company inside the city of Madrid). 
This pilot showcases the ability of our solution to deal with \emph{large dataset conversion}.

In Milano, our \emph{infrastructure manager} pilot, we transformed into NeTEx some data sources of SEA\footnote{SEA Aeroporti, cf. \url{http://www.seamilano.eu/}}, the company managing both Linate and Malpensa airports. In particular, we focused on two different datasets:  the airport facilities description (e.g., help desks, information points, ticket booths, lifts, stairs, entrances and exit locations, parking, gates) and the airport internal transport data (i.e.,  shuttle service between terminals).
This pilot showcases the ability of our solution to build \emph{custom conversion pipelines} that take multiple input sources into account and that involve \emph{proprietary data formats}; indeed, we converted both datasets into an integrated NeTEx representation.

In Genova, our \emph{transport operator} pilot, we involved AMT\footnote{AMT Genova, cf. \url{https://www.amt.genova.it/amt/}}, the public transport service provider of the metropolitan area of Genoa. 
This pilot showcases the ability of our solution to take a generic GTFS-to-NeTEx conversion pipeline (similar to the one used for the Madrid pilot) and to customise it by adding an initial \emph{data preparation} step to operate data cleaning and data enrichment of the original GTFS source provided by the transport operator.

\subsection{Mappings' Definition}\label{sec:mappings}

% 1. *** output always netex, ref ontology is transmodel onto based on transmodel (link) ***
In all the pilots presented above, the expected outcome is data expressed in the NeTEx standard. Since NeTEx is a concrete XML serialization of Transmodel~\cite{netex} and since Transmodel is a rich UML representation of the transportation domain, we took Transmodel as reference conceptualization. Therefore, together with the OEG group of UPM, 
%\footnote{Ontology Engineering Group http://www.oeg-upm.net/}, 
we kick-started an effort to produce the ontological version of Transmodel; even if the ontology is far from being complete, the initial modules that allowed us to  address our pilots' data conversion are already available\footnote{Cf. \url{https://w3id.org/transmodel/}.}. %Transmodel ontologization is out of scope for this paper, however, the interested reader can find the documented core modules at

At the time of writing, the following modules, encompassing 285 classes, 178 object properties and 66 data properties, are available: Organisations (information about the public transport organisations), Fares, Facilities, Journeys (trips and different types of journeys for passengers and vehicles), and Commons. It is important to note that the union of all these modules does not allow representing yet all the information that is expressed in the Transmodel specification, since the current focus was only to enable pilots' data conversion.

% 2. all pipelines involve lifting and lowering
All data conversion pipelines share two steps: the \emph{lifting} from the original data format to the Transmodel ontology and the \emph{lowering} from the ontological version to NeTEx. The former requires the definition of lifting mappings in RML, the latter implies the creation of Apache Velocity templates to generate the final output.

% 3. lifting includes...
To understand the complexity and the difficulties in defining the \textbf{lifting mappings}, we schematize the process in terms of its required steps, as follows:
%Lifting mappings defines how to map data from a generic input format to the correspondent RDF triples modelled through a common reference ontology. The steps characterizing this activity are:
\begin{enumerate}
    \item \emph{Assessing the input data model}, to identify the main concepts and relations  in the input data. %format should be identified.
    \item \emph{Aligning the identified concepts/relations with the respective ones in the reference ontology}, considering both the explicit information and the data that could be inferred. Some special cases should also be considered, in particular, cases where a one-to-one mapping cannot be identified. % e.g. "unrolling" lists. % constant information for the specific input sources -> esempio file con shops in stazione, stazione non rappresentata nei dati ma concetto che posso includere come informazione costante data dal fatto che quel file è associato agli shops in una stazione %Once pointed out the relevant information present in the file, the related classes and properties in the reference ontology should be identified to align the terminology. This activity should consider all information directly represented, but also information that can be inferred or obtained through manipulation of the input data. Moreover, it should consider if some properties can be materialized as constant information for the specific input sources analysed, even if this cannot be directly extracted from data (e.g., if a CSV file contains as rows the list of shops in a Station, I should create an RDF triple for each \emph{shop} materialized linking it to the individual representing the \emph{station}, even if this information is not directly represented in the data).
    \item \emph{Extending the reference ontology}, in case some concepts or properties of the input data are not present in the reference ontology. In particular, this can be required if: (i) an information in the input format cannot be represented by the ontology but cannot be discarded, or (ii) the RDF representation of a given piece of information in the reference ontology requires the instantiation of intermediate entities and relations that cannot be materialized given a lack of information in the input data format.%\notes{non ho capito nemmeno questo esempio -> se per esempio nell'ontologia per collegare A e C devo creare le relazioni A->B e B->C ma nei dati di input non ho informazione su B allora sono costretto ad estendere l'ontologia per creare la relazione A->C se non voglio perdere questa informazione}.
    \item \emph{Coding the actual lifting mappings}, which implies creating the files that encode all the above, as well as specific configurations, such as custom functions applied in the process (e.g., to change the format for date and time). Considering the RML specification, a human readable text-based representation can be defined using YARRRML~\cite{heyvaert2018declarative} then compiled to RML, and the Function Ontology can be used to declare functions in mappings \cite{de2016ontology} (FNO is very useful for conversion of specific formats, like date-time for example: if the source doesn't use \texttt{xsd:dateTime}, a custom conversion function is needed to generate the proper triples).%The last activity is related to the choice of the mapping mechanism for the lifting procedure and its configuration. Considering RML, mappings are encoded through a customized configuration file tailored to access data in the input format and to materialize the proper RDF triples using the reference ontology. To guarantee a richer output or to align data formats (e.g., to change the format for date and time), RML supports the specification of custom functions to manipulate data extracted during the materialization procedure. 
\end{enumerate}
%\textcolor{blue}{forse inizierei già qui a spiegare quali sono le difficoltà e i "costi"...}
% lavoro manuale, complessità degli schemi, iterazione dell'approccio, ...

The definition of lifting mappings from GTFS to Transmodel (Madrid and Genova pilots) required the completion of all the above mentioned steps. The alignment was possible only through a deep study and understanding of the two data models and their complexity, since they are completely different, even if covering overlapping concepts. The need for ontology extension was very limited but the RML mapping coding required the definition of several custom functions to properly access and manipulate the input data. 

In other words, this lifting activity was expensive, because it required a lot of manual work; however, the definition of mappings for a widespread format such as GTFS can be reused with a large set of potential customers, hence this activity cost can be recovered through economy of scale. The definition of lifting mappings for proprietary formats (Milano pilot) required a similar effort, which however cannot be leveraged with additional customers, hence it should be taken into account when valuing such custom data conversion activities.

% 4. lowering includes...
% \subsubsection{Lowering step.}
% The \textbf{lowering mappings} define how to extract the data represented in the Transmodel ontology (by querying the knowledge graph generated during lifting) and how to transform them into the required output. Since all our pilots have NeTEx as output format, we can reuse the lowering mappings across the different scenarios. 
Similarly to the lifting procedure, we schematize the steps of the \textbf{lowering mappings} definition as follows:
%Lowering mappings define how to convert a knowledge graph modelled with the reference ontology to an output format. The steps characterizing this activity are:
\begin{enumerate}
    \item \emph{Assessing the output data model}, to identify the main concepts and relations expected in the output format.
    \item \emph{Aligning with the reference ontology}, to identify how to query the knowledge graph expressed in the reference ontology to get the information required to produce the output format; the querying strategy should take into account both the explicit facts and the information that can be extracted with non trivial queries. Special cases should also be considered, in particular, cases requiring the definition of queries with complex patterns involving multiple entities.%\notes{Informazione costante... ma è importante?} %Once pointed out the information that should be represented in the output format, it must be defined which classes and properties of the reference ontology should be queried to extract the data. This activity should take into account all information directly represented by mean of individuals and properties, but also information that can be inferred using not trivial queries. Moreover, it should consider if some data structure of the output format can be generated as constant information from the specific domain.
    \item \emph{Extending the output format}, in case the knowledge graph contains relevant information that is not foreseen in the output format; of course, this is possible if those extensions are allowed by the output format, otherwise the additional knowledge should be discarded. %In some cases, it is possible that an extension of the output format is required to represent some information in the knowledge graph not directly encodable in the output data format. This type of activity is possible if the output data format allows some kind of flexibility in their structure; otherwise, the additional information would be discarded and not converted.
    \item \emph{Coding the actual lowering mappings}, which implies creating the actual templates that encode all the above to generate the output data. Considering Velocity templates, this coding phase includes: %The last activity is related to the choice of the mapping mechanism for the lowering procedure and its configuration. Considering templates the steps are:
    \begin{inparaenum}[(i)]
        \item identifying the output data structures to compose the skeleton for the Velocity template,
        \item defining the set of SPARQL queries  to extract the relevant data from the knowledge graph,
        \item encoding the  logics to iterate on the SPARQL results and properly populate the skeleton according to the output data format.
    \end{inparaenum}
\end{enumerate}
% 4b. what is difficult/expensive and why
It is worth noting that the choice of Transmodel as reference conceptualization has a twofold advantage: on the one hand, as already said, Transmodel is a broad model that encompasses a rich spectrum of concepts and relations of the transport domain, and, on the other hand, it is the model on which the NeTEx XML serialization was defined, hence the correspondence between the ontology and the output format is quite large.

Therefore, the definition of lowering mappings between the Transmodel ontology and NeTEx was quite straightforward given the strict relation between the two models and the use of similar terminology. For this reason, the assessment of the output data model and the reference ontology alignment activities were simplified, and the output format extension activity was not necessary. Moreover, since all our pilots have NeTEx as output format, we reused the lowering mappings across the different scenarios. 

\subsection{Pipeline Composition}\label{sec:pipeline}

% 1. 3 pilots require different pipelines because of requirements
The three pilots illustrated in Section~\ref{sec:snap} have different requirements, hence they need different data conversion pipelines. In this section, we explain the three pipelines we built on top of Chimera, to show the flexibility of our approach. In general, setting up a pipeline is a matter of adapting and configuring Chimera blocks to address the specific scenario.
%The SNAP solution mainly relies on the Chimera converter described in Section \ref{sec:chimera}. In this section, we showcase the flexibility of the proposed approach in generating customised semantic conversion pipelines for the SNAP pilots.

% 2. basic pipeline --> madrid
The basic pipeline, implemented for the \emph{Madrid pilot}, executes the conversion from GTFS to NeTEx. 
This case demonstrates the main advantages of our solution in composing custom conversion pipelines. The first advantage is the possibility to create custom blocks in Chimera pipelines, besides the default ones. In this case, a custom \emph{GTFS Preprocessing} block checks the file encoding (e.g., to handle UTF with BOM files) and generates input streams for the lifting procedure (e.g., creating different \emph{InputStream}s from a single file to overcome a known limitation of the current RML specification, i.e. the impossibility to filter rows in CSV data sources). 
The second advantage is the possibility of including existing Camel blocks in the pipeline. In this case, the \emph{ZipSplitter} block accesses the different files in the zipped GTFS feed, and other utility blocks deal with input/output management and routing.
To design the final pipeline, the \emph{lifting} and \emph{lowering} blocks are configured with the specific mappings, and the additional blocks are integrated to define the intended flow:
\begin{inparaenum}[(i)]
    \item an \emph{AttachGraph} block to inizialize the connection with the remote RDF repository,
    \item a \emph{ZipSplitter} block,
    \item the custom \emph{GTFS Preprocessing} block,
    \item a \emph{RMLProcessor} lifting block configured using RML mappings from GTFS to the Transmodel Ontology,
    \item a \emph{TemplateProcessor} lowering block configured using a Velocity template querying a knowledge graph described by the Transmodel Ontology and producing NeTEx output, and
    \item a \emph{DumpGraph} block to serialize the content of the generated knowledge graph.
\end{inparaenum}

%\begin{figure}[h!]
%  \centering
%  \includegraphics[width=\linewidth]{fig/SEA.png}
%  \caption{Chimera pipeline customised for SEA. Some blocks are not represented to simplify the diagram.}
%    \label{fig:sea}
%\end{figure}

% 3. adding different data/format --> milano
The pipeline implemented for SEA in the \emph{Milano pilot} extends the Madrid pipeline showcasing how Chimera allows to easily integrate data in different formats. In this scenario, data on airports' facilities provided in a proprietary data format should be merged with GTFS data describing the shuttle service. To produce a unified pipeline for SEA, we configure Chimera to execute in parallel two lifting procedures:
\begin{inparaenum}[(i)]
\item a GTFS-to-Transmodel lifting portion, similar to the Madrid one, to obtain shuttle service data represented in the Transmodel ontology, and 
\item a custom SEA-to-Transmodel lifting portion, to obtain facility data represented in the Transmodel ontology. 
\end{inparaenum}
After the lifting procedures, a \emph{DataEnricher} block is added to the pipeline to merge the RDF triples materialized in the two different lifting procedures. Finally, the Transmodel-to-NeTEx lowering templates are executed to map the shared graph to an integrated NeTEx representation.
%As a result we produced two XML NeTEx output files describing the airports on two different layers an integrating the two input sources. In the \emph{Transport Infrastructure} file we described airports, terminals, levels, entrances, gates, equipment, car parks and the bus stop places. In the \emph{Transport Services} file we described the shuttle service and we create references to physical bus stops described together with other facilities in the first file.
%\notes{Specify liftings can be executed in parallel}

%As represented in Figure \ref{fig:sea}, 
%\begin{figure}[h!]
%  \centering
%  \includegraphics[width=\linewidth]{fig/AMT.png}
%  \caption{Chimera pipeline customised for AMT Genova. Some blocks are not represented to simplify the diagram.}
%    \label{fig:amt}
%\end{figure}

% 4. adding data preparation with external system --> genova
The pipeline implemented for AMT in the \emph{Genova pilot} extends the Madrid pipeline by integrating Chimera with functionalities of external systems. In this case, the conversion pipeline needed to interact with an external data preparation component. This external system, 
%component, 
based on Fiware\footnote{Fiware, cf. \url{https://www.fiware.org/}.},
enriches an initial GTFS feed with additional data sources and it can be configured to call a REST API whenever a new enriched feed is available. In the implemented pipeline, pre-existing Camel blocks are configured to accept POST calls containing an enriched GTFS feed. Each API call triggers the GTFS-to-NeTEx conversion pipeline, as in the other pipelines, and in the end also returns the results to the initial requester.
% as shown in Figure \ref{fig:amt}

% 5. ***why it was easy to customize pipelines with chimera***
The pipelines of the three pilots demonstrate how the Chimera modular approach, inherited from Camel, allows to easily define customized conversion pipelines. The basic conversion pipeline can be configured with  the default blocks for lifting and lowering. Complex scenarios can be addressed implementing additional blocks or employing already defined Camel blocks. Moreover, the same pipeline can be manipulated to fulfill different requirements with minimal modifications.

\subsection{Performance Testing}\label{sec:performances}

%*********************** BEGIN COMMENT ***************************
% 1. evaluation set up + target of "reasonable time/memory" (?)
% 3a. migliorie sul mapper -- abbiamo qualche numero?
% 3b. migliorie sui mapping -- abbiamo qualche numero?
% 3c. migliorie sul lowering -- abbiamo qualche numero?
% 4. *** why our solution is good ***
%*********************** END COMMENT ***************************

In this section, we provide some statistics on the actual execution of the Chimera conversion pipelines presented in Section \ref{sec:pipeline}  and we comment on performance results and addressed issues.

%Performance limitate dalla macchina e dalla versione Free che però sostiene il lowering
We executed the conversion using Docker Containers on a machine running CentOS Linux 7, with Intel Xeon 8-core CPU and 64 GB Memory. Memory constraint is set to 16GB using the Docker \texttt{--memory-limit} option on running containers, no limits are set on CPU usage. GraphDB 9.0.0 Free by Ontotext\footnote{GraphDB, cf. \url{http://graphdb.ontotext.com/documentation/9.0/free/}.} was used as remote RDF repository by Chimera pipelines for storing the materialized knowledge graph and querying triples during the lowering phase.

In Table \ref{tab:stats}, we report the execution results. For each conversion case, we detail: the dataset size (for GTFS feeds, the total number of rows in all CSV files); the number of triples in the materialized knowledge graph; the lifting, lowering and total execution times; the NeTEx output dataset size.

The numbers show that NeTEx is much more verbose than GTFS and, since Transmodel has similar terminology, the knowledge graph and the output dataset are much bigger than the input data. 
% Another aspect related to mappings is that the lifting of some concepts in GTFS generates a greater amount of triples if compared to other concepts. For this reason, a feed A may generate less triples of a feed B, even if the size of the feed B is greater than the one of A.

The conversion times showcase the ability of our solution to handle large knowledge graphs. Considering the  size of CRTM, AMT and EMT datasets, we notice how conversion time grows almost linearly with respect to the input size. The biggest dataset conversion required one hour, which is reasonable for batch processing of static transport data that changes sporadically.
% It is important to point out, since results shown in the table don't allow to consider separately different steps of the pipeline, that time required for lifting is often higher than time required for lowering. 

To make our solution efficient, we implemented in Chimera a set of optimizations as follows. %In the following paragraphs, we report hints and implemented techniques, now available in Chimera, that made efficient the developed solution. Moreover, we explain how the execution time can be easily further optimized.
The rml-mapper library, used by the Chimera lifting block, stores all triples generated in an in-memory object before serializing them at the end of the procedure. To reduce memory consumption, we implemented incremental upload of triples generated to a remote repository during the materialization. This modification reduced the memory consumption, but shifted the performance bottleneck to the triplestore. For this reason, we allowed to set in the pipeline the number of triples in each incremental \emph{insert} query, we created a queue collecting pending queries and we spawn a thread pool of configurable size to consume the queue. Since the Free version of GraphDB allows for only two concurrent queries on a single core, the execution time is affected and directly influenced by the performances of the single core, slowing down the queue consumption.

%\bgroup
%\def\arraystretch{1.1}%  1 is the default, change whatever you need
\begin{table}[t]
%\small
\centering
%\resizebox{.8\textwidth}{!}{
\begin{tabular}{lc|c|c|c|c}
%\hline
 &  & \textbf{CRTM} & \textbf{EMT} & \textbf{SEA} & \textbf{AMT} \\
\hline
\emph{GTFS total no. rows} &     & 168,759    & 2,381,275  & 106     & 710,104  \\ %\hline
\emph{GTFS size} & MB    & 12.46 & 100  & 0.005  & 45 \\ %\hline
\emph{Other data size} & MB & - & - & 1.62*  &  -\\ \hline
\emph{Lifting time} & sec      & 121 & 1,697 & 6 & 546 \\ %\hline
\emph{Lowering time} & sec      & 98 & 1,823 & 5 & 624 \\ %\hline
\emph{Conversion time} & min:sec      & 3:39 & 58:40 & 00:11 & 19:30 \\ \hline
% \emph{Max Memory Usage}      & 3.33 GB & 8.39 GB & 0.6 GB & 6.23 GB \\ \hline 
% Tolto perchè bisognerebbe fare un discorso più ampio considerando il fatto che in realtà il "carico" in termini di memoria è stato spostato sul DB
\emph{No. triples} & & \textcolor{white}{-}1,692,802\textcolor{white}{-}   & \textcolor{white}{-}27,981,607\textcolor{white}{-} & \textcolor{white}{-}7,597\textcolor{white}{-}  & \textcolor{white}{-}9,361,051\textcolor{white}{-}  \\ %\hline
\emph{NeTEx size} & MB & 182.3  &  2,450 & 0.728  & 502 \\ %\hline
\end{tabular}
%}
\\
%\footnotesize{*Only a subset of this dataset (i.e. airports facilities) is converted to NeTEx.}
%\medskip
\caption{Conversion execution statistics; * indicates facilities in Malpensa and Linate airports, only a subset of this dataset is converted to NeTEx.}
%\emph{CRTM}: GTFS feeds of Metro, Light rail and Regional railways in the Madrid area, \emph{EMT}: GTFS feed EMT Madrid, \emph{SEA}: GTFS Navetta Aeroportuale Malpensa and facilities in Malpensa and Linate airports (*only a subset of this dataset is converted to NeTEx), \emph{AMT}: enriched GTFS feed AMT Genova.}
\label{tab:stats}
%\vspace{-4mm}
\end{table}
%\egroup

We also experienced that a clear performance improvement in the RML materialization comes from a reduction of the number of \emph{join} conditions in the RML mappings. In RML, triple generation is often achieved with joins between different data sources; this powerful RML construct, however, is computationally expensive, especially in case of large datasets and nested structures. In our tests, join conditions exponentially increase the required conversion time. Therefore, in our mappings we opted to limit the use of join conditions and instead use separate and simpler Triple Maps with the same IRI generation pattern, to achieve the same result without heavily affecting the conversion time. %even when they could be linked together
%They can be often substituted using fields employed in the \emph{join} condition to generate the same IRI in \emph{triples map} referring different logical sources and, thus, creating the expected relation when the overall graph is composed.

%Considering the lowering procedure based on Apache Velocity templates, an efficient triplestore is strictly required to perform complex queries within a reasonable amount of time. Moreover, the lowering performance is directly related to the query performance, thus, for example, it is important to reduce the number of \emph{optional} statements in SPARQL. Nonetheless, we decided to code our Transmodel-to-NeTEx templates with \emph{optional} patterns, to avoid assumptions on the queried knowledge graph. However, it is possible to get a consistent speedup in the execution time by defining more specific and performant SPARQL queries. 
Considering the lowering procedure, since a standard lowering approach has not emerged yet, we adopted a generic templating solution based on Apache Velocity, which is flexible to adapt to any target format. Any SPARQL SELECT query supported by the triplestore (that returns a table) can be included in the template; the processing of the SPARQL result table can be as complex as needed (using the Velocity Template Language). In the described scenario of Transmodel-to-NeTEx lowering, we implemented a set of generic templates, each with SPARQL queries with multiple \emph{optional} patterns, to avoid assumptions on the available data. Of course, writing templates with  focused queries (e.g. avoiding \emph{optional} clauses) would improve the processing time, because the lowering performance is directly related to the query performance.

To further reduce the execution time, we generated supporting data structures to access data more efficiently, with queries avoiding nested loops. To reduce memory consumption, we removed white spaces and newlines in the Apache Velocity template, negatively impacting the output readability, which can however be improved at a later stage.

\subsection{Business Viability}\label{sec:business} %Change name of the section? %commercial Assessment

Finally, we add some considerations regarding the business evaluation of the presented work.

Initially, the goal of our Chimera framework was limited to provide a \emph{data conversion solution}, to ensure the compliance with the mentioned EU Regulation. This objective was fully achieved, and all the involved stakeholders -- representing transport authorities, transport operators and infrastructure managers -- expressed their full satisfaction with the results obtained with our solution. We run a final meeting with each of them to illustrate the approach, the results, the pricing model and to ask for feedback. They all received their input data enriched, integrated and transformed into the required NeTEx format, with no effort on their side. Regarding the conversion effectiveness, they all agreed that we had successfully provided a solution to the problem of compliance to the European regulation and that they will adopt it, once the respective National Access Point will be setup, requiring them to supply their data in the NeTEx format.

%\textcolor{red}{We did not quantify the stakeholder satisfaction; we run a workshop with each of them to illustrate the approach, the results, the pricing model and to ask for feedback. Regarding the conversion effectiveness, they all agreed that we had successfully provided a solution to the problem of compliance to the European regulation and that they will adopt it, once the respective National Access Point will be setup, requiring them to supply their data in the NeTEx format.}

With respect to the conversion issue, therefore, we showed the feasibility and viability of our solution. Furthermore, to preserve our business proposition, we keep the lifting and lowering mappings together with the pipeline configuration as our unique selling point, while we decided to release the Chimera framework as open source, as already mentioned, to facilitate its adoption and improvement by the larger community.

It is worth noting that developing a conversion solution on top of Chimera still requires all the technical skills to manually develop RML mappings (for the lifting step) and SPARQL queries to fill in the templates (for the lowering step), even when the full domain knowledge is available. This means that there is room for improvement in terms of supporting tooling by our community. At the moment, in business terms, this represents a competitive advantage for us and for Semantic Web-based solution providers.

While addressing the data conversion issue, however, we discovered that we can indeed offer a broader set of services to the target customers in the transport domain. As a result of our business modeling effort, we defined three offers:
\begin{itemize}
    \item \emph{Requirements analysis and compliance assessment}, to investigate the feasibility and complexity of converting the customer data in the desired format;
    \item \emph{Conversion services}, including data preparation and enrichment, lifting to a reference ontology, data integration and lowering to the desired format;
    \item \emph{Supporting services}, including long-term maintenance (e.g., repeated conversion, incremental addition of data sources), knowledge transfer and training.
\end{itemize}

In terms of pricing model, on the basis of the described experience, we now opt for different pricing depending on the \emph{estimated difficulty of the conversion}, which we measure along three dimensions of the source data: (i) popularity of the data format (widespread or proprietary, because this impact the re-usability of the mappings), (ii) data volume, and (iii) input format complexity (again, because it impacts the mapping definition).

As a concluding remark, we highlight that the Madrid transport authority CRTM -- which is already adopting Semantic Web solutions and managing RDF data -- gave us a very positive feedback with respect to our solution's ``collateral effect" of generating a nucleus of a \emph{multimodal transport knowledge graph}: they perfectly understood and appreciated the enabled possibility to build additional added value services on top of the knowledge graph itself, well beyond the pure data conversion for EU Regulation compliance. 
Indeed, once National Access Points will be in place and transport stakeholder will provide their data, the actual exploitation of the multimodal transport knowledge graph will be enabled, e.g. by developing intelligent journey planning offering solutions based on the usage of different transport modes.

The availability of an integrated and harmonized knowledge graph can also pave the way for assessing the conversion completeness. However, in most of the pilot cases illustrated in this paper, the input data sources are in GTFS (a very simple and basic tabular data format), the target standard is NeTEx (a very complex and articulated XML format), and the Transmodel reference ontology is very close to NeTEx. Thus the coverage of the target standard (and the reference ontology) is always quite limited even if the entire information contained in the source data is completely used in the mapping. Therefore we leave this kind of analysis as future work to support and improve the mapping development (e.g. using shape validation to assess the quality/completeness of the mapping results with respect to the reference ontology).

%% file: sections/conclusion.tex
\section{Conclusions and Future Works}\label{sec:conclusion}

In this paper, we presented our solution to enable the conversion of transport data, into standards required by the European Commission, using Semantic Web technologies.
% We motivated our approach discussing the twofold benefit of semantic interoperability to address challenges in this domain: on one hand, it reduces the overall number of mappings required for conversion, on the other hand, it contributes to build a multi-modal transport knowledge graph that can be used to offer added-value services.
To support the implementation, we designed the Chimera framework, 
% made available as open-source software
%\footnote{Chimera, cf. \url{https://github.com/cefriel/chimera}}, 
providing a modular solution to build semantic conversion pipeline configuring a set of pre-defined blocks.
The described solution has been employed within the SNAP project on concrete scenarios and real datasets involving transport stakeholders in Italy and Spain. The different pilots have been presented proposing an evaluation of the solution on different dimensions.
%: mappings, flexibility, performance and business viability.
The performed activities acknowledge the feasibility of the solution on a technological side, the desirability for stakeholders and the business viability of the approach.
% First of all, based on the experience made in the SNAP pilots, we provided a conceptualization of the different steps required to define lifting and lowering mappings to/from the reference ontology. The schematized process offers a practical way to assess the complexity of the mappings' definition, that is often the most time-consuming activity in the definition of the pipeline. 
% Then, we presented a more technical evaluation discussing:
%\begin{inparaenum}[(i)]
%\item the flexibility of the Chimera framework in adapting to different requirements with minimal effort, and 
%\item a performance assessment reporting statistics on the executions of the implemented pipelines and identifying bottlenecks and possible improvements. 
%\end{inparaenum}

We are also adopting Chimera in the ongoing SPRINT project\footnote{Cf. \url{http://sprint-transport.eu/}} with different reference ontologies and source/target standards, to implement and evaluate conversion pipelines in both cases of batch transformation and message translation; we have already reached a clear scalability improvement with respect to our previous ORM-based solution~\cite{carenini2018st4rt2} and we are proving the generalizability of our technological solution beyond the scenario offered in this paper. %We are also contributing to the benchmarking of different Semantic Web-enabled solutions for the GTFS-to-LinkedGTFS lifting.

% FW: Cover other Transmodel-derived standards for dynamic-data (DATEX II and SIRI), reverse mappings}
As future works, we plan to: 
\begin{inparaenum}[(i)]
\item explore additional tools to facilitate the mappings' definition (e.g. collaborative and visual tools), 
\item implement additional blocks for Chimera to offer more options in the pipeline definition (e.g., approaches based on a virtualized graph to extract data from the original sources) and, 
\item perform an evaluation of the proposed solution considering dynamic data (e.g., DATEX II and SIRI formats) and requirements in real-time scenarios.
\end{inparaenum}

%%%%%%%%%%%%%%%%%%%% Other Notes %%%%%%%%%%%%%%%%%%%%%%%%%%%%%
%The conversion problem is not simple and its requirements can be very different from case to case. In particular, data can be too large for implementing materialisation, i.e., the approach used in this first implementation of Chimera. When the input of the conversion problem is a large dataset in a batch-like scenario, or when a message conversion requires accessing large datasets it may be necessary to adopt a different approach based on virtualization techniques. These methods leverage OBDA (Ontology-Based Data Access) of heterogeneous data sources, also called OBDI (Ontology-Based Data Integration). This approach requires anyway a set of lifting mappings from the input data format to an ontology but it avoids materializing an RDF graph to query it in the lowering phase. Virtualization techniques are based on the assumption that, given the set of lifting mappings, it is possible to rewrite a SPARQL query in a set of queries directly applicable to the format of the input data (e.g. SQL queries), to convert in RDF only the result sets and to merge them to obtain the correct answer. In this way, the RDF graph is not materialized but it is virtually mapped on the original sources.